\documentclass[10pt]{article}
\usepackage{amsmath,amsfonts,amssymb,amsthm,bbm}

\usepackage{cite}
\usepackage{graphicx}
\usepackage[latin1]{inputenc}
\usepackage{amsthm}
\usepackage{hyperref}
\usepackage{color}

\newcommand{\Ref}[1]{(\ref{#1})}

\newcommand{\C}{{\mathbb C}}
\newcommand{\N}{{\mathbb N}}
\newcommand{\R}{{\mathbb R}}
\newcommand{\Z}{{\mathbb Z}}

\newcommand{\cD}{{\mathcal D}}

\newcommand{\SU}{\mathrm{SU}}

\newcommand{\SL}{\mathrm{SL}}

\newcommand{\su}{{\mathfrak{su}}}

\renewcommand{\u}{{\mathfrak u}}

\newcommand{\be}{\begin{equation}}
\newcommand{\ee}{\end{equation}}
\newcommand{\bea}{\begin{eqnarray}}
\newcommand{\eea}{\end{eqnarray}}

\newcommand{\bit}{\begin{itemize}}
\newcommand{\eit}{\end{itemize}}

\newcommand{\tr}{{\rm Tr}}
\newcommand{\f}{\frac}
\newcommand{\tl}{\tilde}
\def\p{\partial}

\newcommand{\Id}{\mathbbm{1}}

\newcommand{\re}{\mathrm{Re}}
\newcommand{\im}{\mathrm{Im}}

\newcommand{\ra}{\rangle}

\newcommand{\bra}[1]{\langle {#1}|}
\newcommand{\ket}[1]{|{#1}\rangle}

\newcommand{\vet} [2] {\left ( \begin{array}{c}{#1}\\{#2} \end{array} \right ) }
\newcommand{\mat} [4] {\left ( \begin{array}{cc}{#1}&{#2}\\{#3}&{#4} \end{array} \right ) }

\renewcommand{\a}{\alpha} \renewcommand{\b}{\beta} \newcommand{\g}{\gamma}  
\renewcommand{\d}{\delta}  \newcommand{\eps}{\epsilon} \newcommand{\z}{\zeta}
 \renewcommand{\th}{\theta}   
  \renewcommand{\k}{\kappa}  \renewcommand{\l}{\lambda}
        \let\om=\omega
 \newcommand{\s}{\sigma}  \renewcommand{\t}{\tau}   \let\vphi=\varphi 
\let\G=\Gamma

\def\T{\mathbbm T}
\newcommand{\norm}[1]{\bra{#1}#1\ra}
\newcommand{\po}{[\pi\ket{\om}}
\newcommand{\tpo}{[\tl\pi\ket{\tl\om}}
\def\ST{{\mathbbm S}^0_\g}
\newcommand{\roc}[2]{[{#1}\ket{#2}}

\newcommand{\bk}[2]{\bra{#1}{#2}\ra}

\begin{document}

\title{Twisted geometries, twistors, and conformal transformations}

\author{{Miklos L$\mathring{\mathrm{a}}$ngvik$^{\dagger,\, \ddagger}$ and Simone 
Speziale$^{\dagger}$}
\smallskip \\ 
\small{$^{\dagger}$ Centre de Physique Th\'{e}orique, CNRS UMR 7332, AMU \& USTV, Luminy 
Case 907, 13288 Marseille, France} \\
\small{$^{\ddagger}$ Department of Physics, University of Helsinki,
P.O. Box 64, FIN-00014 Helsinki, Finland}}
\date{\today}

\maketitle

\begin{abstract}
The twisted geometries of spin network states are described by \emph{simple} twistors, isomorphic to null twistors with a 
time-like direction singled out. The isomorphism depends on the Immirzi parameter $\g$, and reduces to the 
identity for $\g=\infty$. Using this twistorial representation we study the action of the conformal group SU(2,2) on the classical
phase space of loop quantum gravity, described by twisted geometry. The generators of translations and 
conformal boosts do not preserve the geometric structure, whereas the dilatation generator does. It 
corresponds to a 1-parameter family of embeddings of T*SL(2,C) in twistor space, and its action 
preserves the intrinsic geometry while changing the extrinsic one -- that is the boosts among 
polyhedra.
 We discuss the implication of this action from a dynamical point of view, and compare it with a discretisation of the dilatation generator of the continuum phase space, given by the Lie derivative of the group character. At leading order in the continuum limit, the latter reproduces the same transformation of the extrinsic geometry, while also rescaling the areas and volumes and preserving the angles associated with the intrinsic geometry. Away from the continuum limit its action has an interesting nonlinear structure, but is in general incompatible with the closure constraint needed for the geometric interpretation.
As a side result, we compute the precise relation between the extrinsic geometry used in twisted geometries and the one defined in the gauge-invariant parametrization by Dittrich and Ryan, and show that the secondary simplicity constraints they posited coincide with those dynamically derived in the toy model of \cite{IoFabio}.
\end{abstract}

\section{Introduction}

While Loop Quantum Gravity (LQG) (for a recent monograph, see \cite{Rovelli:2014ssa}) is a background-independent 
approach to quantum gravity, the internal Minkowski metric plays a key role in identifying the local gauge group of 
the theory, $\SL(2,\C)$ in the covariant formulation, and SU(2) through the use of Ashtekar-Barbero variables. The Minkowski metric has also a group $\SU(2,2)$ of (four-fold cover of) conformal isometries. In this paper we study the action of this group on the phase space of loop quantum gravity on a fixed graph. This space, that corresponds to the kinematical semiclassical description of the theory, has been shown to be described by a collection of polyhedra \cite{twigeo,IoCarloGraph,EteraSU2UN,IoPoly}, which defines a discrete metric structure with intrinsic and extrinsic components, called twisted geometry. These can in turn be parametrised in terms of twistors, and it is the Hamiltonian action of $\SU(2,2)$ on twistor space that we use for our study.

The interest in studying conformal transformations on this space is two-fold. On the one hand, 
conformal transformations play an important role in classical general relativity, and it would be very 
useful to have any of their applications available in the study of loop quantum gravity, such as identifying the splitting between causal  
structure and conformal factor of the metric, or the behaviour of transition amplitudes under dilatations. 
As twisted geometries are discrete, we may expect difficulties in realising conformal transformations. Indeed, we know from results in Regge calculus that one can not realise in a 
discrete setting conformal transformations in the usual sense of rescaling distances while preserving 
angles \cite{Rocek:1982tj}, ultimately for the simple reason that angles are determined themselves by 
edge lengths and/or areas. As we show in this paper, using twisted geometries and the twistorial representation of the conformal group allows one to go partially beyond this result: while the generators of translations and conformal boosts are not compatible with the constraints present in twisted geometries, the generator of dilatations is, and can thus be meaningfully realised on the phase space. Its geometric 
interpretation is however counterintuitive: first, the 3d intrinsic geometry is invariant, including areas and volumes; second, only the extrinsic geometry changes, the orbits giving linear shifts of the boosts among the polyhedra. This unconventional behaviour for a dilatation may appear puzzling, but it is a 
natural consequence of the relation between the variables of loop quantum gravity and the Lorentz 
algebra. In fact, the intrinsic geometry of loop quantum gravity is built out of angular momentum 
operators, and these are invariant under the dilatations of their embedding conformal group. It is 
only the extrinsic geometry, built out of the holonomies, that can be affected by these dilatations. 
Since changing the extrinsic geometry without changing the intrinsic one affects the spacetime 
curvature, our result suggests that the $\SU(2,2)$ dilatation generator can play a dynamical role in 
the theory, something we briefly comment upon.
It is then natural to ask whether a more conventional dilatation generator exists on the phase space of 
twisted geometries. The situation can be compared with a more geometric notion of dilatations, 
rescaling distances while preserving angles, such as symplectic dilatations in the phase space of 
connections and triads of the continuum theory. While dilatations in this sense do not exist on the 
phase space of a fixed graph because it has compact directions, nor can they be expected from the 
Regge calculus analysis, it is possible to define a discrete version of the continuum symplectic 
dilatations. The nonlinearity of its action can be organised in an interesting way, and its effect 
on the geometry studied explicitly. At leading order in the continuum limit, the latter reproduces 
the same transformation of the extrinsic geometry, while also rescaling the areas and volumes and 
preserving the angles associated with the intrinsic geometry. However, away from the continuum limit 
its action is highly nonlinear, and in particular, is in general incompatible with the closure 
constraint needed for the geometric interpretation.

The second line of interest in our question is more formal.
In their initial formulation by Penrose, spin networks were conceived to describe only the conformally 
invariant part of a quantum spacetime, via the angles. To introduce distances, he envisaged an 
extension to the Poincare group, or since the latter is not semi-simple, to the conformal group of 
Minkowski spacetime. This program famously led to twistor theory \cite{Penrose72}, based on the double cover of the 
conformal symmetry group of Minkowski spacetime SO(4,2), or its double cover SU(2,2).
Later on, SU(2) spin networks found a key role in loop quantum gravity, where they provide an orthonormal basis of the theory. In their use in quantum gravity, the norm of the angular momentum is interpreted as an area eigenvalue, thus introducing a notion of scale and distances per se. 
Extensions to $\SL(2,\C)$ are also commonly used in linking loop quantum gravity to the spin foam formalism for transition amplitudes.
It is nonetheless still an open and intriguing question to develop Penrose's original program and 
establish a precise relation between SU(2,2) spin networks and the SU(2) ones used in LQG. Such SU(2,2) spin networks would be based on twistors in a similar way as the SU(2) ones are based on spinors, and to understand their relation precisely, it is useful to first clarify how the
conformal symmetry of twistor space is broken in the way loop quantum gravity uses this space. We answer here this question at the classical level, showing that it is the area-matching constraint, responsible for reducing the twistor phase space to the $\SL(2,\C)$ phase space, that partially breaks the conformal symmetry, with only the action of dilatations remaining. Our results will be then useful in future works concerned with understanding the relation at the quantum level.

The paper is organised as follows. In Section 2 we provide a short overview of the link between twistor 
theory and loop quantum gravity. This allows us to introduce the representation of the conformal 
group that we will use later on, but also to provide an overview of the state of the art of twisted 
geometries, that has seen many remarkable developments in the last couple of years.
In Section 3 we show what the linear, primary simplicity constraints used in spin foam models for loop quantum gravity \cite{EPRL} mean 
from the point of view of twistor theory. The Lorentz-invariant part ('diagonal simplicity') selects 
twistors with helicity linked to the Lorentz Casimirs, a constraint breaking full conformal invariance. 
While the helicity is non-vanishing, the constraint structure makes these twistors isomorphic to null 
twistors. The isomorphism depends on the Immirzi parameter $\g$ (reducing to the identity for 
$\g=\infty$), and we refer to these as $\g$-null twistors. The non Lorentz-invariant part 
('off-diagonal simplicity'), depending on a  gauge-fixing time direction, selects a special point along 
the $\g$-null twistor's null ray intersecting the time direction. The resulting \emph{simple} twistors 
(in the sense that they satisfy the simplicity constraints of loop quantum gravity) identify a unique 
space-like plane, that in the geometric reconstruction is used to define the faces of the polyhedra of 
twisted geometries. In Section 4 we compute the action of the conformal group on twisted geometries.
Of the non-$\SL(2,\C)$ generators of $\SU(2,2)$, only the dilatation is compatible with the various 
geometric constraints (area-matching, simplicity and closure). Its orbits preserve the fluxes, and 
change only the holonomy. At the SU(2) level, the change is a linear shift of the  twist angle; this 
transformation thus leaves unaffected the whole 3d geometry, and only changes the 
embedding of the SU(2) holonomy in the covariant phase space by a linear shift. At the $\SL(2,\C)$ level, again the 
transformation only affects the covariant version of the twist angle, which now carries the 
interpretation of boosts among adjacent polyhedra, that is of extrinsic curvature. Defining the 
extrinsic curvature in this way requires fixing the time gauge, and it is desirable to have a 
completely gauge-independent test of this action. To that end, we consider the alternative 
description of extrinsic geometry by Dittrich and Ryan \cite{DittrichRyan}, here extended to 
Lorentzian signature. The latter is fully gauge-invariant, but depends on a choice of edge per 
face of the polyhedron, and gives a unique boost only for shape-matched configurations. We show 
that also the extrinsic curvature defined in this way transforms in the same way under a $\SU(2,2)$ dilatation. As a side result, we work out the exact relation between the two definitions of boost dihedral angles, and we prove that they coincide when certain secondary simplicity constraints are satisfied. Such secondary constraints were derived dynamically using a toy model in \cite{IoFabio}, and our analysis here shows also that they precisely coincide with (the Lorentzian extension of) those defined in \cite{DittrichRyan}. 

Finally, in Section 5 we compare these transformations with those induced by a direct discretisation of the continuum connection-triad symplectic dilatations, obtaining the results anticipated above. Our discretisation takes the form of a Lie derivative of the group character, and when interpreted in terms of spinors, it acts as a boost in mixing the source and target spinors. In spite of its simplicity, its action on holonomies and fluxes is completely nonlinear, and has no resemblance with usual dilatations. This is unavoidable, as the discrete phase space has compact directions, so there is no usual meaning of linear dilatations. On the other hand, the generator has the property of preserving the symplectic structure, and all transformed quantities recover the expected behaviour in the continuum limit.

All our results, computing the action of $\SU(2,2)$ dilatations, the relation between the different boost dihedral angles in the literature and proving the equivalence of different secondary simplicity constraints, as well as computing the finite action of the discrete holonomy-flux dilatation, rely heavily on the use of twistorial formalism, and  are a demonstration of its usefulness to loop quantum gravity.

We use mostly-plus convention for the Minkowski metric and indices $I=0,1,2,3$. $A, B=0,1$ are spinorial indices and $\alpha=1,2,3,4$ twistorial indices.

\section{Twistors and twisted geometries}

A twistor can be described as a pair of spinors, 
$Z^\a=(\omega^A,i \bar\pi_{\dot A})\in\mathbb{C}^2\oplus\bar{\mathbb{C}}^2{}^\ast=:\mathbb{T}$. 
It has a dual, $\bar Z_\a=(-i\pi_A, \bar\om^{\dot A})$, 
which defines a pseudo-Hermitian norm of signature $(++--)$,
\be
Z^\a \bar Z_\a = 2 \im(\pi_A\om^A),
\ee
preserved by SU(2,2) transformations. It is well-known \cite{Penrose72} that these transformations can be realized by Hamiltonian vector fields, if we equip the space with 
canonical Poisson brackets,
\be\label{ZZbar}
\big\{Z^\a, \bar Z_\b\big\}=i\d^\a_\b.
\ee
In Penrose's abstract index convention (see Appendix B), the generators of SU(2,2) can be written as 
\be\label{generators}
J^{IJ} = \om^{(A}\pi^{B)}\eps^{\dot A\dot B} + {\rm cc},
\qquad P^I = i \pi^A \bar\pi^{\dot A}, \qquad C^I = i \om^A \bar\om^{\dot A}, \qquad 
D= \re(\pi_A\om^A),
\ee
and their finite action on a phase space function $f=f(\om^A, \pi^A)$ is realized via the exponential map,
\be
\exp(\epsilon_{IJ} M^{IJ} )\triangleright f :=\exp(\epsilon_{IJ}\{M^{IJ}, \cdot\}) f = 
f+\sum_{n=1}^\infty \frac{1}{n!}\{\epsilon_{IJ}M^{IJ},...,\{\epsilon_{KL}M^{KL},f\}\},
\ee
where $\{\epsilon_{IJ}M^{IJ},...,\{\epsilon_{KL}M^{KL},f\}\}$ is the $n-$fold nested bracket of the generator $M^{IJ}$ with $f$. Half the twistor norm, $s:=\im(\pi_A\om^A)$, is called helicity of the twistor,\footnote{In the twistor literature, the helicity is usually given by the real part. The difference comes from the extra $i$ used here in the definition of the twistor. In this way we can match Dirac's conventions for a bi-spinor, and bridge more easily with the notation used in loop quantum gravity. Notice also that we use metric conventions with mostly plus, thus we map vectors to anti-hermitian matrices, see Appendix A.} 
since 
\be
W^I=- sP^I,
\ee
where $W^I:=1/2\eps^{IJKL} P_J J_{KL}$ 
is the Pauli-Lubanski vector. 
The complex scalar $\pi_A\om^A= D+i s$, for which we also use the index-free notation $\po$, is invariant under the 
Lorentz subalgebra. See Appendix B for more details on the index-free notation for
spinors and their dual.

The representation of the conformal algebra constructed in this way is not the most general one, since the algebra has 15 dimensions, whereas the carrying space $\T$ has only 8 dimensions. In fact, the generators \Ref{generators} are not independent: $P^I$ and $C^I$ are null and related to $D$ and the Lorentz generators by 
\be
2 P^{[I} C^{J]} = \left(D\Id +s\star\right) J^{IJ}, \qquad P^IC_I = -(D^2+s^2),
\ee
where $\star$ is the Hodge-star. Accordingly, the three Casimir operators of $\su(2,2)$ (see the Appendix for definitions) are not 
independent: they are all proportional to the only conformally invariant quantity in $\T$, the helicity:
\be\label{casT}
\mathcal{C}^{(2)} = 6\, s^2, \qquad \mathcal{C}^{(3)} = -6\, s^3, \qquad \mathcal{C}^{(4)} = 6\, s^4.
\ee
Hence, the representation of the algebra on twistor space is special, and furthermore reducible, with irreducibles labelled by $s$. In this sense, Penrose refers to twistors as the spinors of the conformal group.

Irreducible representations with independent real values of the Casimirs can be obtained working with the larger carrying space $\T^2$ made of pairs of twistors $(Z,\tl Z)$, with generators constructed by linearity. This space has now 4 conformally invariant quantities, given by the two helicities plus the pseudo-hermitian product $\bar Z_\a \tl Z^\a$, 
and the Casimirs are independent functions of these four variables. $\T^2$ has been considered in the twistor literature \cite{Tod76, BetteT2}, among other things in relation to ambitwistor theory and to build representations of massive particles. 
But as shown in \cite{IoWolfgang}, $\T^2$ has another interesting property that links it to loop quantum gravity: it contains $T^*\SL(2,\C)$ as a symplectic submanifold,  obtained by Hamiltonian reduction with respect to the first class constraint
\be\label{defC}
C := \po - \tpo =0,
\ee
referred to as (complex) area-matching condition.
Explicitly,  we take
\be\label{hfdef}
J^{IJ}  = \om^{(A}\pi^{B)}\eps^{\dot A\dot B}+cc, \qquad
h^A{}_B = \f{\tl\om^A \pi_B - \tl\pi^A\om_B}{\sqrt{\po}\sqrt{\tpo}}, \qquad \tl J^{IJ}  = -\tl\om^{(A}\tl\pi^{B)}\eps^{\dot A\dot B}+cc,
\ee
and
following the conventions of \cite{IoWolfgang}, we take opposite signs for the twistor brackets\footnote{This is convenient for the geometric interpretation of the theory. Switching to the alternative conventions with equal-sign brackets also used in the literature is straightforward, via the map $\om^A\mapsto \pi^A$, $\pi^A\mapsto -\om^A$.} :
\be \{\pi_A, \om^B\} = \d_A^B = - \{\tl\pi_A, \tl\om^B\} \ee
It is easy to check that \Ref{hfdef} are $C$-invariant. When the constraint \Ref{defC} is satisfied, the generators are 
related by the adjoint action of $h$ and thus \Ref{hfdef} span a 12-d submanifold of $\T^2$, for which 
one recovers the Poisson brackets of $T^*\SL(2,\C)$ with $J$ and $\tl J$ acting respectively as left-invariant and right-invariant vectors fields on the group manifold, parametrized by $h$ as a left-handed group element. This symplectic reduction holds provided $\po\neq 0$, which we exclude from our analysis from now on, and is 2-to-1: there is a $\Z^2$ symmetry associated with the simultaneous sign flip of the spinors. See \cite{IoWolfgang,WielandTwistors,IoTwistorNet} for more details.

The symplectic manifold $T^*\SL(2,\C)$ is the building block of projected spin networks \cite{EteraProj}, which appear as boundary states of covariant spin foam models \cite{DupuisLivine,IoCarloCov}. There, a copy of $T^*\SL(2,\C)$ decorates each link of an oriented graph, and the orientation is used to associate unambiguously $Z$ to the source node, and $\tl Z$ to the target. 
Each twistor is then subject to the primary simplicity constraints, which reduce the covariant boundary states to those relevant for the SU(2) spin networks of loop quantum gravity. 
The constraints are deduced from a discretisation of the gravitational action in Ashtekar variables, and relate the tetrad to the bivector $B$ which is the field canonically conjugated to the connection. 
To represent  the constraints at the discrete level, we introduce a time direction $N^I(n)$ on each node $n$ of the graph.
In their linear form introduced in \cite{EPR} (see also \cite{EPRL,FK,LS2}) the primary simplicity constraints read
\be\label{simplCov}
N_I B^{IJ} = 0,
\ee
for all bivectors $B$ associated with links sharing the node $n$,
and imply simplicity of $B$, namely $\eps_{IJKL}B^{IJ}B^{KL}=0$.
The bivector $B$ is related to the Lorentz generators via $B = (\Id -\g\star) J$, where $\g \in \mathbb{R}$ is the Immirzi parameter. See \cite{EPRL,IoHolo,Wieland1} for details, and \cite{IoNull} for extensions to the case of a null hypersurface.

It is customary the fix the time gauge  $N^I=(1,0,0,0)$, 
but the construction extends to an arbitrary gauge \cite{IoHolo}.  In this gauge, we identify the left-handed generators with 3-vectors $\Pi^i$, whose real and imaginary parts are rotations $L^i$ and boosts $K^i$, via $\Pi^i=(L^i+i K^i)/2$. 
The normal $N^I$ is conserved by the canonical SU(2) subgroup of the Lorentz group, and allows us to define the Hermitian scalar product between spinors, $\bra{\om}\pi\ra:=\delta^{A \dot A}\pi_A\bar{\omega}_{\dot A}$. We will also use the short-hand notation $\|\om\|^2:=\norm{\om}$ for the norm. 

The constraints \Ref{simplCov} read $K^i+\g L^i=0$, and are equivalent to imposing the matching of left-handed and right-handed sectors up to a phase, 
\be\label{simpl}
\Pi^i = -e^{i \th}\bar\Pi^i, \qquad \g=\cot\f\th 2.
\ee
Using \Ref{generators}, the constraints have simple spinorial equivalents \cite{WielandTwistors,IoWolfgang},
\begin{align}\label{Fdef}
F_1=D-\g s=0, 
\qquad
F_2= \bra{\om}\pi\ra=0.
\end{align}
The constraint $F_1$, real and Lorentz-invariant, is solved posing
\be\label{po}
\po=(\g+i)j, \qquad j\in \R
\ee 
and fixes the relative global phase between the spinors.
$F_2$, which explicitly depends on the chosen time direction, implies that one spinor is 
proportional to the parity transform of the other, $\pi_A \propto  \d_{A\dot A}\bar \omega^{\dot A}$. Putting the two conditions together, the constraints can be solved expressing one spinor in terms of the other and the norm $j$:
\be\label{pi=om}
\pi_A = \frac{(\g+i)j}{\|\omega\|^{2}} \d_{A\dot A}\bar \omega^{\dot A}.
\ee

There is an alternative parametrization of the solution space, motivated by the canonical structure of the constraints: while $F_2$ is second class, $F_1$ is first class. A good coordinate among its orbits is the SU(2) norm $\|\om\|^2$, and the $F_1$-gauge-invariant spinor is
\be\label{defz}
z^A := \sqrt{2j}\frac{\omega^A}{\|\omega\|^{i\g+1}}, \qquad \|z\| = \sqrt{2j}.
\ee
Here $j\neq 0$ by assumption, and we have restricted to $j>0$ using the $\Z^2$ symmetry of the symplectic reduction by $C$.
See  \cite{IoWolfgang} for details. 

The solution space is thus a five-dimensional manifold $\ST$, dependent on the Immirzi parameter (via $\th$) and the choice of time gauge, that can be parametrized by one of the two Lorentzian spinors, plus the norm $j$; or alternatively, by the reduced spinor $z^A$, plus the norm $\|\om\|$ :
\be\label{simpletwi}
\ST = \{ \om^A, j\} = \{z^A, \|\om\|\}.
\ee
The two norms $j$ and $\|\om\|$ play complementary roles in the geometric interpretation of loop quantum gravity : $j$ is the area of the face dual to the link, while $||\om||$ is related to the extrinsic curvature, as we will recall below. 
We refer to twistors satisfying the simplicity constraints \Ref{Fdef}, or equivalently \Ref{pi=om}, as \emph{simple} twistors.\footnote{The term simple twistors is also used in the twistor literature, but in reference to a bi-twistor being simple in the same sense of a simple bivector as defined above.}

On a link, we impose the same set of constraints on both source and target twistors,
\be\label{zred}
\f{z^A}{\|z\|} = \frac{\om^A}{\|\om\|^{i\g+1}}, 
\qquad  \f{\tl z^A}{\|\tl z\|} = \frac{\tl\om^A}{\|\tl\om\|^{i\g+1}},
\ee
with induced Poisson brackets $\{z^A, \bar{z}^{\bar B}\} = i \d^{A\dot B} = - \{\tl z^A, \bar{\tl z}^{\bar B}\}$,
and reduced area matching condition $C=\|z\|^2-\|\tl{z}\|^2=0$. The latter implies $\|z\|^2=\|\tl{z}\|^2= 2j$, and reduces this $\C^4$ symplectic space to $T^*\SU(2)$, as shown in \cite{twigeo2}, parametrized by
\be\label{T*SU2}
\vec X = \f12\bra{z}\vec\s\ket{z}, \qquad 
 g = \f{\ket{\tl z}\bra{z}+|\tl z][z|}{\|z\| \|\tl z\|},
\qquad \vec {\tl X} = -\f12\bra{\tl z}\vec\s\ket{\tl z},
\ee 
where $\vec\s$ is the vector of Pauli matrices and $X^2 = \tl X^2 = j^2$.
Again, $X$ and $\tl X$ are related on-shell of $C$ by the adjoint action of $g$, and act as right-invariant and left-invariant vector fields.

That is, twistors constrained by the area matching and primary simplicity constraints describe the phase space $T^*\SU(2)$ of loop quantum gravity. 
When SU(2) gauge-invariance holds, by means of the closure conditions $\sum_{l\in n} X_l = 0$ at each node of the graph,\footnote{Here $\tl X_l$ is to be used if $l$ is incoming instead of outgoing from $n$.} the space describes twisted 
geometries, a generalization of Regge geometries with polyhedra glued along faces that have unique areas but a priori 
different shapes \cite{twigeo,IoPoly,DittrichSpeziale}. The intrinsic geometry is defined from the shapes of the polyhedra, 
reconstructed from the $X_l$ vectors interpreted as normals to the faces \cite{IoPoly}. The extrinsic geometry between the 
polyhedra can be defined via a Lorentzian dihedral angle, that is the boost relating the four-dimensional normals between 
adjacent polyhedra  \cite{IoWolfgang}:\footnote{Alternatively, it can be defined via the edge vectors as done in 
\cite{DittrichRyan}. The precise relation and on-shell equivalence of these two constructions will be proved below in 
Section \ref{SecBiancaAngle}.}
\be
\cosh\Xi := -\tl N_I \Lambda^I{}_J(h) N_J,
\ee
where $\Lambda(h)$ is the Lorentz transformation induced by
the parallel transport $h$ in the vectorial representation. On shell of the simplicity constraints, this gives
\be\label{defXi}
\Xi = \ln \f{\|\om\|^2}{\|\tl\om\|^2}.
\ee

Notice that at first sight the extrinsic geometry, described by the boost $\Xi$, is missing in the reduced SU(2) phase \Ref{T*SU2}. However, this is not the case: the reduced holonomy carries information about boosts.
To understand this subtle but crucial point, notice first that the $\SU(2)$ obtained from the reduction is \emph{not} (the double cover of) the rotation subgroup of $\SL(2,\C)$, but is instead embedded in $T^*\SL(2,\C)$ in a non-trivial way, which in particular spans also the boost directions. 
To understand this, let us look at the value on shell of the simplicity constraints of the left-handed connection, which is given by
\be\label{honshell}
h = \f{ e^{-(1+i\g)\f\Xi2} \ket{\tl z} \bra{z} + e^{(1+i\g)\f\Xi2} |\tl z] [z|}{\|z\| \|\tl z\|}.
\ee
This expression shows that the reduced SU(2) is related to the left-handed SU(2) in a $\Xi$-dependent way. Recall that in the continuum, the Ashtekar-Barbero connection is defined as
$A^i = -\f12\eps^{ijk} \om^{jk} + \g\om^{0i}$, where $\om^{IJ}$ is the Lorentz connection, and its boost part $\om^{0i}$ 
gives the extrinsic curvature. Here $i=1,2,3$ is a spatial index in the internal space.
In terms of the left-handed part $\om^i:=(\Id+i\star)\om^{0i}$, this can be rewritten as $\om^i = (1+i\g)\om^{0i} -iA^i$. Equation \Ref{honshell} is precisely a discrete version of this relation, and shows that the reduced holonomy $g$ is indeed a discrete version of the Ashtekar-Barbero connection.\footnote{To avoid confusion, let us stress that it is not the holonomy of the Ashtekar-Barbero connection: that wold require imposing the simplicity constraints at every point of the path, as suggested in \cite{AlexandrovNewVertex}, but that is a procedure not available if one works, as in here, with a discrete phase space associated with a fixed graph. 
}

The above mixing of rotations and boosts, central to the use of Ashtekar-Barbero variables, is the consequence of the $\g$-phase in the simplicity constraints \Ref{simpl}, or equivalently in the spinors \Ref{defz}.
It shows that the SU(2) discrete data do carry information on the extrinsic geometry, namely the Lorentz boost $\Xi$. In order to extract this information however, like in the continuum, one needs the embedding of the SU(2) phase space in the covariant one. The latter is provided by the secondary simplicity constraints, which express $\om^{ij}=\om^{ij}(E)$, and thus allow to read the extrinsic curvature from the reduced data, $\om^{0i}= \om^{0i}(A,E)$.
As discussed in \cite{IoWolfgang} and showed explicitly in \cite{IoFabio}, the discrete equivalent of this mechanism 
goes as follows: the secondary simplicity constraints form a second class pair with $F_1$, and thus imposing them gives 
an explicit gauge fixing of the orbits of $F_1$. This gauge fixing provides
a function $\Xi_l=\Xi_(\{z_l\})$ (typically non-local on the graph),
that is a way to express the extrinsic geometry in terms of the
reduced SU(2) data, as desired.

To explicitly write the secondary constraints, and provide explicit formulas for this procedure, it is 
best to work in terms of SU(2) gauge-invariant quantities. A natural and familiar basis for these are 
the Wilson loops $\tr \prod_{l\in f}h_l$ on the graph. Using spinors, an alternative convenient basis 
is provided by the set over all nodes of the products $\bk{z_i}{z_j}$ and $[z_i\ket{z_j}$ of spinors in
the half-links $i,j$ sharing a node (In the rest of this section we will not use internal
indices $i$ anymore, so we use $i, j$ etc to refer to half-links).
This basis,\footnote{Notice that both bases are redundant, the first by the Mandelstam identities 
associated with the recoupling theory \cite{Rovelli:1991zi},
the second by the Pl\"ucker relations $[z_i\ket{z_j}[z_k\ket{z_l} = [z_i\ket{z_l}[z_k\ket{z_j} + [z_i\ket{z_k}[z_j\ket{z_l}$. 
These latter have the further advantage of generating an algebra, 
with subalgebra $\u(N)$ generated by  the $ \bra{z_i}z_j\ra$ alone.}  introduced in \cite{EteraSU2UN} for loop quantum gravity, can be given a geometric 
parametrization as follows  \cite{FreidelJeff13}, 
\begin{align}\label{z-geo}
& \bra{z_i}z_j\ra = \sqrt{4 j_i j_j} \cos\f{\phi_{ij}}2 e^{-\f i2 (\a^i_j - \a^j_i)}, \qquad
 [z_i\ket{z_j} = \eps_{ij} \sqrt{4 j_i j_j} \sin\f{\phi_{ij}}2 e^{\f i2 (\a^i_j + \a^j_i)}.
\end{align}
Here we assumed that both links $i$ and $j$ are outgoing from the node,\footnote{If one is incoming, the required parity map is explicitly given by $P\ket{z}=-|z]$.} and $\eps_{ij}=\pm$ is needed for the phases $\a^i_j$ to be defined in $[0,2\pi)$.
The geometric interpretation of these invariants relies on a map from the graph to its dual cellular decomposition, such that that each link of the graph is dual to a face which is shared by two polyhedra dual to the nodes connected by the link. 
Then, recall through \Ref{T*SU2} that each spinor defines a vector. Then $j_i$ is the norm of this vector, and $\phi_{ij}$ the angle between the vectors $\vec X_i$ and $\vec X_j$.
When the closure constraint at a node is satisfied, the vectors define a unique convex and bounded 
flat polyhedron, whose shape and adjacency matrix can be reconstructed using for instance the algorithm
given in \cite{IoPoly}.\footnote{See also \cite{HaggardPentaVolume} for analytic formulas for the 
5-valent case.}  Then for adjacent faces $(ij)$, $\phi_{ij}$ are the 3-dimensional external dihedral 
angles, $\vec X_i\times \vec X_j$ the edge vector, and taking scalar products between the edge vectors, 
one can immediately check that $\a^i_{jk}:=\a^i_j-\a^i_k$ are the 2d angles between the vectors 
associated with the edges $(ij)$ and $(ik)$ and thus, when they are adjacent, the 2d dihedral angles of the 
polygon. \footnote{The phases $\a^i_j$ also admit a direct geometric interpretation in terms of the framing 
vector of \cite{EteraSU2UN}, but we will not need it here.
Notice also that this geometric reconstruction is actually 2-to-1 for the $\Z_2$ symmetry discussed 
earlier, a direct consequence of the usual extra sign appearing when describing a spinor in terms 
of its flagpole and flag. 
}

In the usual twisted geometry parametrization, one extracts the direction of the normal vectors via the Hopf section 
$$n=n(\z)=\f1{\sqrt{1+|\z|^2}} \mat{1}{\z}{-\bar\z}{1}: S^2\mapsto \SU(2),$$ where $\z:={z}^0/{z}^1$ and similarly for the tilded spinors with $\tl\z:=\tl z^0/\tl z^1$. Then, \Ref{T*SU2} can be rewritten as
\be\label{twigeos}
\vec X = 2j \tr(n\t_3n^{-1}\vec\t), \qquad g= \tl n e^{\xi\t_3} n^{-1}, \qquad 
\vec {\tl X} = -2 j \tr(\tl n\t_3\tl n^{-1}\vec\t), 
\ee
where $\vec{\tau}= - i/2 \vec \sigma$ and $\xi:=2\arg \tl z^1 - 2\arg z^1 \in[0,4\pi)$ parametrizes the remaining freedom in $g$ at fixed $X$ and $\tl X$.
As the directions and the norms are used to reconstruct the metric properties of the individual 
polyhedra, $\xi$ is the natural candidate to describe the extrinsic geometry among them, which as we 
argued above, is in covariant terms described by the boost $\Xi$. However, there are two non-trivial 
aspects concerning the exact relation between $\xi$ and $\Xi$: (i) $\xi$ is not SU(2)-gauge-invariant, 
while $\Xi$ is; (ii) $\xi$ is an SU(2) angle, while $\Xi$ is a Lorentzian boost. The first issue was 
initially dealt with simply by assuming to work in a fixed gauge \cite{twigeo,IoCarloGraph}. 
An alternative proposed in \cite{DittrichRyan} is to consider a gauge-invariant definition of the twist angle via the scalar product between the edge vectors associated to the same edge in the two frames of the source and target nodes:
\be
\cos\xi^i_{jk}:= \f{(\vec X_{\tl \imath}\times \vec X_k) \cdot g_i\triangleright (\vec X_i\times \vec X_j)}{|X_{\tl\imath}\times X_k| \, |X_i\times X_j|}.
\ee
The price to pay for explicit gauge-invariance is the edge dependence: only for shape-matched configurations the above angle is independent of the choice of edge. 
Thanks to the spinorial gauge-invariant basis \Ref{z-geo}, it is also possible to compute the SU(2)-gauge invariant twist angle \cite{FreidelJeff13}
as follows,
\be
\xi^i_{jk}:=\a^{\tl\imath}_{k}- \a^i_{j} 
= \xi_i + \vphi^{+-}_{ik} - \vphi^{--}_{ik} - \vphi^{+-}_{ij} + \vphi^{--}_{ij} + \f\pi2(\eps_{ik}-\eps_{ij}),
\ee
where $\vphi^{AB}_{ij}$ is the phase of the matrix elements $\bra{A}n^\dagger(\z_i) n(\z_j)\ket{B}$ in the canonical basis.
The above definition is orientation-dependent, and to fix ideas we have taken a link $i$ with $j$ outgoing from its source 
node and $k$ outgoing from its target node.
We see that analogously to the Wilson loop trace, this gauge invariant quantity mixes the twist angles $\xi_i$'s and the 
directions $\z_i$'s, as it should; it has the advantage of doing so in a way that a single $\xi_i$ enters instead of 
all of those belonging to the loop. The price to pay is that the definition is not unique: 
for each link $i$, the gauge-invariant twist angle depends on a choice of a pair of links $(jk)$ connected to the source and 
the target of $i$. That is, from the point of view of a triangulation dual to the graph, for each triangle, it depends on a 
choice of edge.
Notice that this is the same redundancy that appears in the construction of Dittrich and Ryan 
\cite{DittrichRyan}, that will be discussed in Section \ref{SecBiancaAngle}.\footnote{The redundancy can be fixed choosing one representative per link, for instance the average as suggested in \cite{DittrichRyan}. 
Once the choice is made, together with the areas $j_l$ and the shape variables for each polyhedron 
(2 for tetrahedra, $2f-6$ for a general polyhedron with $f$ faces), one obtains reduced variables for 
the gauge-invariant phase space.}
In terms of these variables it is also immediate to characterise the shape-matching conditions: these are given by
\be\label{shapematching}
\a^i_{jk} = \a^{\tl\imath}_{lm} \qquad \Leftrightarrow \qquad \xi^i_{jk} = \xi^i_{lm},
\ee
namely the matching of the 2d dihedral angles \cite{DittrichSpeziale} in the first form, or the requirement of a unique gauge-invariant twist angle per link in the second form.

Then, point (ii) is dealt with precisely by the secondary constraints, which relate the extrinsic curvature $\Xi$ to the SU(2) data. Secondary simplicity constraints have been proposed by Dittrich and Ryan \cite{DittrichRyan,DittrichRyan2} as a direct discretisation of the Levi-Civita condition $\om^{ij}=\om^{ij}(E)$, but also derived as the stabilisation of the primary simplicity constraints in a toy model where a discrete Hamiltonian constraint could be fully solved \cite{IoFabio}. As we show below in Section \ref{SecBiancaAngle}, the two definitions coincide, and read
\be\label{gXixi}
\g\Xi_i - \xi^i_{jk} = 0.
\ee
This formula provides the sought relation between the extrinsic curvature and the reduced SU(2) phase space, and is the state of the art of the understanding of twisted geometries. 
Notice that these secondary constraints imply the shape-matching conditions \Ref{shapematching} \cite{DittrichRyan,DittrichRyan2,IoFabio}.\footnote{An alternative definition of discrete secondary constraints has been proposed in \cite{HaggardSpin}, which does not require the shape-matching conditions. It is however gauge-dependent, and the relation of the extrinsic geometry on the SU(2) data via this alternative construction has not been computed yet, to the best of our knowledge.}

To complete this overview of twisted geometries, let us briefly comment on dynamical aspects. The main interest in 
twisted geometries comes from their relation to spin foam amplitudes and loop quantum gravity, for which they describe 
the kinematical semiclassical limit on a fixed graph \cite{twigeo}. One can then study the dynamics induced by the quantum 
theory, and this is a non-trivial task with still a number of
open questions. In particular, a key result \cite{BarrettLorAsymp} is that in the large spin limit of the EPRL 
(Engle-Pereira-Rovelli-Livine) model on a 4-simplex, the shape matching conditions are satisfied and the amplitude is dominated by the Regge action. 
This is an encouraging result for the model, however the limiting procedure is delicate to handle on a 
full triangulation, and it has been argued that only flat solutions are compatible with the saddle 
point equations of the large spin limit \cite{HellmannFlatness}. This has started a debate in the 
literature on whether non-trivial curvature is properly accounted for, and if not, what is the problem 
with the model or the limiting procedure. See \cite{Bonzom:2009hw,AlexandrovSimplClosure,RovelliMagic} 
for some references. Furthermore, the result relies on the special geometric properties of the 
4-simplex, and it is not clear how to interpret the most general spin foams that are not dual to a 
triangulations \cite{KKL, Ding:2010fw}. These partial results show the importance of improving our 
understanding of the dynamics already at the classical level. Actions for twisted geometries and their 
relations to the Regge action on-shell of the shape matching conditions have been studied in 
\cite{DittrichSpeziale,FreidelJeff13,WielandNew,DittrichRyan,EteraSpinor,HanZhangLor}, but there is as
of yet no clear consensus on the meaning of curvature and dynamics away from the shape matching subspace, 
nor on the off-shell role of the torsion a priori kinematically present in the theory \cite{IoWolfgang}. An interesting 
development in this sense is the alternative interpretation of the same phase space in terms of discrete geometries with 
torsion along the edges proposed in \cite{FreidelSpinning}. 
Studying the dynamics of the non-shape-matched configurations is also important to understand whether they admit a continuum limit reproducing general relativity, or whether the latter property can be satisfied only by the Regge configurations.
Lastly, spinors have also found many applications in the study of the quantum dynamics of spin foam models, e.g.
\cite{DupuisLivineHolomorphic1, DupuisLivineHolomorphic2, IoHolo, Freidel, Hnybida1, Hnybida2, 
BonzomCostantinoLivineIsing}.

\section{Geometry of $\g$-null and simple twistors}

Before studying the action of conformal transformations on twisted geometries, let us look at what the simplicity 
constraints imply on the usual geometric picture of a twistor.
The latter is derived from the incidence relation
\be\label{incidence}
\om^A = i X^{A\dot{A}}\bar \pi_{\dot A},
\ee
solving it for $X^{A\dot{A}}(\om,\pi)$ and interpreting the result as a curve in Minkowski spacetime.\footnote{In this section, we follow Penrose's notation 
and use $X^{A\dot A}=X^I$ to refer to a point or a family of points in Minkowski
space as derived from the incidence relation. However, the temptation of identifying its spatial part with the fluxes $\vec X$ previously defined should be resisted. As this section shows, when the simplicity constraints are satisfied,  $X^{A\dot A}$ is
proportional to the time-like direction of the constraints, and the
fluxes $\vec X$ are the spatial parts of the spinor's null poles (the null vector associated with the spinor).} 
When the twistor is null, namely $s=0$, $X^{A\dot{A}}$ is anti-hermitian and the solution of \Ref{incidence} is
\be\label{nullray}
X^{A\dot{A}}=-\f{i \om^A \bar\om^{\dot{A}}}{{\po}} + b \, i \pi^A\bar\pi^{\dot{A}}, \qquad b\in \R.
\ee
It describes a null ray  in the direction of the null-pole of $\pi^A$, that is $i \pi^A\bar\pi^{\dot{A}}$, going through a point at a distance $-1/\po$ 
from the origin along the null-pole of $\om^A$, that is $i \om^A \bar\om^{\dot{A}}$. The subspace of null twistors is denoted $\N$.

For non-null twistors, the general solution of \Ref{incidence} is instead
\be\label{a-plane}
X^{A\dot{A}}=-\f{ i \om^A \bar\om^{\dot{A}}}{{\po}} + \l^A\bar\pi^{\dot{A}}, \qquad \l^A\in \C^2.
\ee
This gives a 2-plane in complexified Minkowski spacetime, spanned by the tangent vectors $\l^A\bar\pi^{\dot{A}}$, 
called $\a-$plane: it is totally null, and its associate bi-vector is self-dual. Its only intersection with real 
Minkowski space is along the null ray $i\pi^A\bar\pi^{\dot A}$. However, it is also possible to give an alternative 
interpretation of  non-null twistors in real Minkowski spacetime, by looking at the set of all null twistors $Y^\a$ 
incident with $Z^\a$, 
\be
s(Y)=0, \quad \bar Z\cdot Y=0.
\ee
This set defines a three-parameter family of null rays associated with the twistor $Z^\a$.
A classic explicit evaluation \cite{Penrose72} identifies this family with the Robinson congruence, 
twisting to the right (\emph{righty}) or twisting to the left (\emph{lefty}) according to the sign of $s$, thus 
giving the name twistor to $Z^\a$.

Notice that all three geometric pictures (null-ray, $\a-$plane, Robinson congruence) are invariant under a complex 
rescaling of the twistor, 
\be\label{rescalings}
(\om,\bar\pi)\mapsto \l(\om,\bar\pi), \qquad \l\in\C,
\ee
thus more precisely they give a geometric representation of projective twistor space ${\mathbbm P}\T$.
On the other hand, in the LQG interpretation the twistor's norms $j$ provide the values of the areas of the 
triangulation, so it is truly $\T$ and not ${\mathbbm P}\T$ that one works with. Conversely, the sign of the 
helicity does not matter: both righty and lefty twistors give the same geometry, and restriction can be made to 
positive helicity $j>0$.

Let us now come to the geometric interpretation of the simple twistors solutions of the simplicity constraints 
\Ref{Fdef}. From \Ref{po} we have that $s=j$, thus since we are avoiding the degenerate configurations $j=0$, the 
simple twistors are not null. As we can always restrict to $j>0$, we see that they they describe right-handed 
Robinson congruences. However, the phase shift
\be\label{pig}
(\om^A, \pi_A) \mapsto (\om^A,  
e^{-i\th/2}{\pi}^{A})
\ee
maps a twistor satisfying \Ref{po} to a null twistor: the space $\N_\g$ of solutions to $F_1=0$ is isomorphic to $\N$.
The isomorphism reduces to the identity at $\th=0$ or $4\pi$, that is $\g=\pm\infty$: Proper null twistors could only 
play a role  when the simplicity constraints are $L^i=0$, that is in the limit where the Engle-Pereira-Rovelli-Livine (EPRL) model \cite{EPRL} reduces 
to the Barrett-Crane \cite{BC} model.
Accordingly, we may refer to twistors satisfying $F_1=0$ as ``$\g-$null'', and associate a null ray to them via
\be\label{Xg}
X_\g^{A\dot{A}}=-\f{ i \om^A \bar\om^{\dot{A}}}{j\sqrt{1+\g^2}} + b \, i \pi^A\bar\pi^{\dot{A}}, \qquad b\in \R.
\ee
Then, the restriction imposed by $F_2=0$ is to make the null pole of $\pi^A$ proportional to the null pole of $\om^A$.
Further, the light ray passes through the point
\be
X_\g^I = -\f{\sqrt{2}||\om||^2}{j\sqrt{1+\g^2}} N^I,
\ee
consistently with \Ref{pi=om}, and this fixes
\be
b=-\f{||\om||^4}{j^3(1+\g^2)^{3/2}} .
\ee
That is, the condition \Ref{pi=om} is picking up a point from the incidence relation, that corresponds to the time-like vector $N^I$ specified by the time gauge. Hence, simple twistors describe a subspace of $\N_\g$ with a time-like vector singled out. 
Finally, notice also that 
\be
\arg \om^0 +\arg\om^1 +\arg\pi^0+\arg\pi^1 = \th
\ee
so the planes of the spinor's flag are related by a rotation of an angle $\pi-\th$ in the space-like plane. Clearly, 
the construction generalises from the time gauge $N^I=(1,0,0,0)$ considered so far to an arbitrary timelike vector
introduced by the simplicity constraints \eqref{simplCov}.

Summarizing, a simple twistor is a $\g-$null twistor with a fixed point along the null ray, so as to align its null 
poles to the timelike vector provided by the
time gauge, and the flag of $\om$ identifies a spacelike bivector normal
to the time direction. The procedure picks up a unique spacelike
plane, the one orthogonal to the spatial parts of the null poles, or,
equivalently, to $L^i$. As discussed below \eqref{simpl} this plane is identified,
in covariant terms, by the simple bivector
$
B \propto (\Id-\g\star)J,
$
and indeed it can be checked that this is the case simply plugging the solution to the simplicity constraints 
in the twistor definition of $J$ \Ref{generators}.
Hence, the association of a twistor with a light ray becomes secondary upon imposition of the 
simplicity constraints, and attention is instead given to this space-like plane: It is indeed this 
plane, the building block of twisted geometries, that defines the face of a polyhedron, with the norm $j$ 
fixing its area.

\section{Twisted geometries and conformal transformations}

The existence of a representation of the conformal group plays an important role in Penrose's vision of 
twistor space. In fact, the data is used to describe spinning massless particles, which are themselves conformally 
covariant. The generators acquire a physical interpretation, for instance $P^I$ is the energy-momentum of the particle, 
and $s$ the helicity. On the other hand, loop quantum gravity suggests an alternative geometric interpretation of 
twistors as twisted geometries, in which the constraints \eqref{defC} and \eqref{Fdef} plus the closure constraint imposing SU(2) gauge 
invariance at the nodes, reduces twistor space to that of twisted geometry. Therefore, the generators of the conformal group, and 
the transformations they induce, can acquire a new geometric interpretation. To that end,
the action of the group should be compatible with the three sets of constraints defining twisted geometries, that is area matching, simplicity and closure. 
A first glance immediately shows that full conformal covariance is broken, as already the (real part of the) area-matching condition imposes $D=\tl D$, which is not preserved by translations and special conformal transformations,
\be
\{D, P^I \} = -P^I, \qquad \{D, C^I \} =C^I.
\ee
Hence, these generators do not preserve the constraint surface, and their action does not admit an interpretation in terms of holonomies and fluxes.
On the other hand, the dilatation generator commutes with all the constraints:
it commutes with $C$, and furthermore 
\be\label{DJ}
\{D, J^{IJ}\}=0,
\ee 
thus it commutes also with the simplicity and closure constraints. 
The GL$(2,\C)$ subgroup generated by $J^{IJ}$ and $D$ is thus compatible with all the constraints. 

Unlike in applications of twistor theory to solving the wave equations, the breaking of conformal covariance here is not introduced by the infinity twistor $I_{\a\b}$, which would preserve the isometry group at infinity (Poincar\'e or (A)dS), but rather by the fixing of the dilatation generators of the two twistors, and preserves instead the subgroup generated by Lorentz transformations and dilatations. 

Therefore, orbits of $D$ live in the phase space of loop quantum gravity and can be given a geometric interpretation.
On each half-link, say the one associated with the source node, $D$ generates symplectic dilatations on the twistor phase space,
\be\label{Daction}
\exp(\l D)\triangleright \ket{\om} = e^{\l/2} \ket{\om}, \qquad \exp(\l D)\triangleright  
\ket{\pi} = e^{- \l/2} \ket{\pi}, \qquad \lambda \in \mathbb{R}.
\ee
On each link, we can consider arbitrary real linear combinations of the generators at source and target,
$\cD:= \l D + \tl \l \tl D.$
The action of $\cal D$ on the holonomy-flux algebra gives
\begin{subequations}\label{Dhf}
\begin{align}\label{DPi}
& \exp(\cD)\triangleright \Pi = \Pi, \qquad \exp(\cD)\triangleright  \tl\Pi = \tl\Pi, \\
& \exp(\cD)\triangleright  h = 
\f{e^{-u/2} \ket{\tl\om}[\pi| - e^{u/2} \ket{\tl\pi}[\om| }{\sqrt{\po} \sqrt{\tpo}} = 
 \cosh(u/2) h + \sinh(u/2) \hat h,  \label{finD}
\end{align}\end{subequations}
where $u:=\l+\tl\l$ and
\be
\hat h^A{}_B  := \f{\tl\om^A \pi_B + \tl\pi^A\om_B}{\sqrt{\po}\sqrt{\tpo}} = 
-\f 4{\po} (h\Pi)^A{}_B,  \qquad \det \hat h=-1.
\ee
It is easy to check that the transformed group element still has unit  determinant for all real values of $u$,
thus the orbits span a one parameter family of embeddings of $\SL(2,\C)$ in $\T^2$. 
This symmetry can be understood as follows: when we embed $T^*\SL(2,C)$ in twistor space, see \Ref{hfdef}, the group element is constructed to satisfy
\be
h \ket{\om} = \ket{\tl\om}, \qquad h \ket{\pi} = \ket{\tl\pi},
\ee
and sends an orthogonal basis into another orthogonal one. 
However, any pair of spinors provides an orthogonal basis provided $\po\neq0$, without restriction on the norms, since there is no Lorentz-invariant norm in spinor space. Therefore, one could have equally well demanded parallel transport of the basis up to a rescaling, that is
\be
h^u \ket{\om} = e^{-u/2} \ket{\tl\om}, \qquad h^u \ket{\pi} = e^{u/2} \ket{\tl\pi}.
\ee
The solution to this equation is precisely \Ref{finD}, and it is clear that any such parametrization would provide an equally good embedding of $T^*\SL(2,\C)$ in twistor space. The action of $D$ on the phase space corresponds to this symmetry of the symplectic reduction from twistor space.

Next, we would like to understand the geometric meaning of this transformation.
First of all, we already know from \Ref{DJ}, or equivalently \Ref{DPi}, that the fluxes are unchanged. 
Indeed, $J^{IJ} = \om^{(A} \pi^{B)}+cc$ is invariant under the symplectic dilatations \Ref{Daction} 
generated by $D$. However, the Casimirs of the Lorentz generators determine the scales in loop quantum 
gravity. Therefore, the symplectic dilatations generated by $D$ do not act as geometric dilatations, as 
they preserve all properties of the three dimensional intrinsic geometry determined by the fluxes: the 2d 
and 3d angles of the polyhedra, but also their areas and volumes which would rescale under a geometric 
dilatation.

So what is the effect of \Ref{Dhf} on the geometry?
Since $\tl \Pi = -h\Pi h^{-1}$, the holonomy is determined by the fluxes up to a rotation and a 
boost in the direction of $\Pi$. These two variables have been introduced in \cite{twigeo2} as 
the twist angle $\xi$ and the boost $\Xi$ and  form an almost abelian pair with the Lorentz 
Casimirs, 
\be
\{D,\Xi\} = 1, \qquad \{D, \xi\}=\g, \qquad \{s,\Xi\}=0, \qquad \{s,\xi\}=1.
\ee
From the first two brackets we immediately read the effect of a finite transformation on each link:
\be\label{DXi}
e^{{\cal D}}\triangleright  \Xi = \Xi + u, \qquad e^{{\cal D}}\triangleright  \xi = \xi + \g u.
\ee
This is the effect of ${\cal D}$ on twisted geometries: it shifts the boost and twist angle among adjacent polyhedra.

To interpret these transformations, let us first look at $\xi$ and the reduced action on $T^*SU(2)$. We will then discuss the meaning of changing $\Xi$, and prove that the same interpretation holds if we abandon the time gauge and use instead the gauge-invariant parametrization of Dittrich and Ryan \cite{DittrichRyan}.

\subsection{Action of $D$ on the reduced phase space and preservation of shapes}

On the SU(2) phase space, the reduced dilatation when the constraints \eqref{Fdef} are imposed reads $D = \g j$, and its action is
\be
\{ D, z^A\} = -\f i2\g z^A, \qquad e^{\l D}z^A = e^{-\f i2\g\l}z^A.
\ee
Together with the rotation generators $X^i$, $D$ generates a $U(2)$ action on the spinors, 
already considered in the LQG context for instance in \cite{EteraSpinor}. 
As for the full $\SL(2,\C)$ phase space, the transformation induced by $D$ does not change the fluxes, but only the holonomy, 
\begin{subequations}\begin{align}
& \exp(\cD)\triangleright  \vec X = \vec X, \qquad \exp(\cD)\triangleright  {\vec{\tl X}} = {\vec{\tl X}} \\
& \exp(\cD)\triangleright  g = \f{e^{\f i2\g u} \ket{\tl z}\bra{z}+ e^{-\f i2\g u}|\tl z][z|}{\|z\| \|\tl z\|}
= \cos\left(\f\g2 u\right) g + i \sin\left(\f\g2u\right) \hat g,\label{Dg}
\end{align}\end{subequations}
where as before $u:=\l+\tl\l$, and 
\be
\hat g = \f{\ket{\tl z}\bra{z} - |\tl z][z|}{\|z\| \|\tl z\|}, \qquad \det \hat g = -1,
\ee
while $\det(\exp(\cD)\triangleright  g )=1$. Again, the orbits represent a symmetry of embeddings of $T^*\SU(2)$ in $\C^4$, which is now represented by defining $g$ as the parallel transport of an orthogonal basis up to a global phase.
Using the Hopf sections and the usual twisted geometry parametrization, it is easy to see that the transformation of $g$ is simply the one induced by the transformation of $\xi$ in \Ref{DXi}.
However, this shift leaves completely invariant the 3d geometry: it only affects the relation between $\xi$ and $\Xi$, that is, the way $g$ is embedded in $T^*\SL(2,\C)$. This can be hinted at by the limit case $\g=0$, for which $g$ is a pure rotation and $D$ vanishes on shell of the simplicity constraints thus leaving $g$ invariant. To make it more explicit, observe that on the full space space $\xi$ is given by 
\be
\xi = 2\arg \tl z^1 - 2\arg z^1= 2\arg \tl \om^1 - 2\arg\om^1+\g \Xi.
\ee
That is, the transformation \Ref{DXi} of $\xi$ under $D$ is the one induced by the transformation of $\Xi$. 
In other words, $g$ transforms under $D$ because it is not a pure 3-dimensional holonomy, but rather an Ashtekar-Barbero holonomy mixing rotations and boosts.

To complete the argument that the 3d geometry does not change, let us verify the effect of $D$ on the SU(2) gauge-invariant basis \Ref{z-geo}. We consider the dilatation generated by an arbitrary linear combination on the graph,
\be\label{Dgraph}
\cD_\Gamma:= \sum_l (\l_l D_l + \tl \l_l \tl D_l).
\ee 
This gives
\begin{align}
& \exp(\cD_\Gamma)\triangleright   \bra{z_i}z_j\ra  = e^{\f i2\g(\l_i-\l_j)} \bra{z_i}z_j\ra, \qquad
 \exp(\cD_\Gamma)\triangleright   [z_i\ket{z_j} = e^{-\f i2\g(\l_j+\l_i)} [z_i\ket{z_j}.
\end{align}
From these we deduce that for a finite dilatation areas and 3d dihedral angles do not change, whereas 
the spinor phases transform as 
$
\d \a^i_j = -\g\l_i.
$
Notice that the transformation is independent of the lower label $j$ in $\alpha_j^i$.
As a consequence,
\begin{align}\label{dxigi}
& \d \a^i_{jk} = 0, \qquad
 \d \xi^i_{jk} = \g (\l_i + \tl \l_i).
\end{align}
The 2d-dihedral angles $ \a^i_{jk}$ are unchanged, consistently with the fact that dilatations do not affect the fluxes. 
The gauge-invariant twist angle $\xi^i_{jk}$ is affected, consistently with the holonomy transformation \Ref{Dg}. 
However, it is changed in a way that is independent of the lower labels, thus the transformation can not change 
whether the shapes match or not. 
Furthermore, from \Ref{DXi} and \Ref{dxigi} we see that the transformation preserves the secondary constraints \Ref{gXixi}.

We conclude that the action of ${\cal D}$ does not change the intrinsic 3d geometry. In particular, it does not break the embedding provided by the secondary constraints, nor the shape matching conditions. These properties make it compatible also as a transformation on the Regge phase space. 

\subsection{Unfixing the time gauge}\label{SecBiancaAngle}

The way we parametrized the covariant twisted geometries, in particular the boost dihedral angle $\Xi$, uses explicitly the time gauge. This is simply a convenience to use the primary simplicity constraints in the linear form \Ref{simpl}, and all results can be shown to be covariant under change of gauge (e.g. \cite{IoTwistorNet}). However, one can also consider a formulation of the theory in terms of purely gauge-invariant quantities and quadratic simplicity constraints. In this setting, we can not use a definition such as \Ref{defXi} to define the extrinsic curvature, but we can define it via the edge vectors of the triangulation. This construction was proposed and carried out in \cite{DittrichRyan} for euclidean signature. Here we extend it to Lorentzian signature, and because the construction is based on the self-dual/antiself-dual splitting, it means dealing with complex vectors instead of real vectors. We then show that the resulting angle is transformed by $D$ in exactly the same way as $\Xi$. In doing so, we work out the explicit relation between the dihedral angle of \cite{DittrichRyan}  and $\Xi$, and find out that the secondary simplicity constraints proposed in \cite{DittrichRyan} are precisely the ones arising in \cite{IoFabio}.

To fix ideas, consider again an oriented graph $i$ dual to a triangulation, with $j$ outgoing from its source node and $k$ outgoing from its target node. We define the anti-selfdual edge vectors
\be
\vec Y_{ij} := 
\vec\Pi_{i} \times \vec\Pi_{j}.
\ee
The same edge is shared by the adjacent tetrahedron, with associated edge vector $\vec Y_{ik}$. The fact that there are two or more independent edge vectors associated with the same edge is just another way of seeing the familiar shape mismatching of twisted geometries. Nonetheless, we can define a (complex) angle among these two vectors as follows,
\be\label{defThB}
\cosh \th^i_{jk} = \f{\vec Y_{\tl \imath k} \cdot h_i \triangleright \vec Y_{ij}}{\sqrt{\vec Y_{\tl \imath k}^2} \, \sqrt{\vec Y_{ij}^2}}.
\ee
This formula adapts the definition of Dittrich and Ryan \cite{DittrichRyan} to Lorentzian signature, using complex angles as in the set-up of Sorkin  for Lorentzian triangulations \cite{Sorkin:1975ah}. Notice that this definition associates three angles to each face (in the case of a triangulation, or as many as the valence of the face in the case of a more general cellular decomposition). 
The shape mismatch is then characterised by the fact that these angles take different values, and the shape matching 
conditions read $\th^i_{jk}= \th^i_{lm}$. We thus have exactly the same redundancy and structure of the shape-matching 
conditions as we do when using reduced SU(2) variables and twisted geometries; see \Ref{shapematching}.

To study the action of ${\cal D}$ on these angles, we first have to parametrize them in terms of Lorentzian spinors. 
Nicely, one obtains a simple expression with only two terms,
\be\label{BJspinors}
\cosh \th^i_{jk} = \f12 
\f{\roc{\om_i}{\pi_j} \, \roc{\om_i}{\om_j} \, \roc{\tl\pi_i}{\om_k} \, \roc{\tl\pi_i}{\pi_k}
+  \roc{\pi_i}{\om_j} \, \roc{\pi_i}{\pi_j} \, \roc{\tl\om_i}{\pi_k} \, \roc{\tl\om_i}{\om_k}}
{\sqrt{\roc{\om_i}{\om_j} \roc{ \pi_i}{\pi_j} \roc{\om_i}{\pi_j} \roc{\pi_i}{\om_j}} \sqrt{\roc{\om_k}{\tl\om_i} \roc{ \pi_k}{\tl\pi_i} \roc{\om_k}{\tl\pi_i} \roc{\pi_k}{\tl\om_i}}}.
\ee
The action of the graph generator ${\cal D}$ can then be easily computed from \Ref{Daction}, to give
\be\label{Dtheta}
e^{\cal D}\triangleright \cosh\th^i_{jk} = \cosh(\th^i_{jk} + u_i).
\ee
Here we used the fact that \Ref{BJspinors} with a relative minus sign gives $\sinh  \th^i_{jk}$.
Remarkably, we have found precisely the same result as for $\Xi_i$, see \Ref{DXi}: the dilatation shifts the extrinsic curvature.
Again, the preservation of the shapes follows from the fact that the change on $\th^i_{jk}$ is independent of the edge used to define it.

The result can be made more transparent if we reintroduce the time gauge, and write \Ref{BJspinors} on-shell of the simplicity constraints.
A straightforward calculation then gives
\be\label{cool}
\cosh\th^i_{jk} = \cosh \Big(\Xi_i + i(\g\Xi_i-\xi^i_{jk})\Big).
\ee
This formula gives the relation in the time gauge between the Dittrich-Ryan dihedral angle $\th^i_{jk}$ ($\SL(2,\C)$-gauge-invariant but edge-dependent) and the twisted geometry dihedral angle $\Xi_i$ (edge-independent but not $\SL(2,\C)$-gauge-invariant, only SU(2)).
Using it, we can also prove that the secondary simplicity constraints defined in \cite{DittrichRyan} as
\be
\th^i_{jk} = \bar \th^i_{jk}
\ee
amount precisely to \Ref{gXixi}, which were derived dynamically through a toy model \cite{IoFabio}.
And finally, \Ref{Dtheta} can be derived directly from \Ref{DXi}, and the fact already remarked that the secondary constraints are preserved.

\subsection{Some dynamical considerations}

We have seen that the dilatation generator acts as a translation on the boost between two polyhedra, be it defined via the time gauge as in \Ref{defXi}, or via the edge vectors as in \Ref{defThB}. In a spacetime picture \`a la Regge, this plays the role of the dihedral angle providing the extrinsic geometry of the embedding in a Lorentzian discrete metric. Therefore, the rescaling \Ref{DXi} generated by $D$ changes the extrinsic geometry, and as the 3d geometry is unchanged, it will affect the spacetime geometry. This suggests that $D$ could play a role in dynamical models, something that we leave for future work.
In fact, not all orbits of $D$ are compatible with a discrete dynamics  \`a la Regge. Restrictions for instance arise from the requirement of discretising spacetime with flat 4-simplices. As simple examples of this, we can consider a triangulation with 5-tetrahedra bounding a 4-simplex: in this case it is clear that imposing flatness in the bulk kills any action of $D$, since the 4-dimensional dihedral angles are then uniquely determined by the areas (up to special configurations).\footnote{The action of $D$ would be admissible if one works with a 4-simplex of constant curvature, since we know from the Schl\"afli identity that on a curved 4-simplex a shift of the dihedral angles at constant areas and 3d geometry induces a shift of the 4-volume:
$\sum_l A_l \d\Xi_l = 3\k \d V,$
where $\k$ is the 4-simplex curvature. However, for this interpretation to make sense, the boundary tetrahedra and triangles should be constantly curved themselves, and this is not the usual interpretation of twisted geometries.
Since curved 4-simplices are expected to arise in dynamical loop quantum gravity models with a 
cosmological constant, e.g. \cite{Bahr:2009qc, Han}, it may be that $D$ could play a role in dynamical models with a cosmological constant.} 
Or, consider a triangulation with three 4-simplices sharing a face in the bulk, so that 4d curvature is allowed via the deficit angle associated to the bulk face. By definition of the deficit angle, different curvatures can be distinguished by the different values of the boundary dihedral angles. Thus, it is now possible to vary the boundary dihedral angles at fixed areas, using ${\cal D}$, and induce in this way a different curvature at the bulk face, giving a clear connection between ${\cal D}$ and dynamical aspects. Notice however that the admissible orbits in this case are spanned by a single-parameter family and not by independent shifts.

Generally speaking, a relation between dilatations and dynamics is to be expected, since we know that in the continuum theory the generator of symplectic dilatations in the phase space of triads and connections appears in the Hamiltonian constraint \cite{Thiemann:1996aw}.\footnote{More recently its role has been put to evidence for the dynamics of cosmological models with non-zero $\Lambda$ \cite{FreidelConfLambda}.}
It is then interesting to also consider a discretised version of the connection-triad symplectic dilatation, and compare its action with the one of $D$, the twistor symplectic dilatation.
This is what we do in the next Section.

\section{Holonomy-flux symplectic `dilatations'}

On the reduced phase space $T^*\SU(2)$, the twistor symplectic dilatation $D\approx \g j$ preserves 
areas and volumes. The scale-invariance of the transformation is also clear if we look at the continuum 
limit of $D$: this gives the norm of the Ashtekar flux $E^i$, which commutes with all fluxes and thus 
the full 3-dimensional metric. 
At the continuum level, a symplectic dilatation in the phase space of Ashtekar-Barbero variables is 
generated by
\be\label{Wcont}
W = E^a_i A_a^i, \qquad \exp(\l W)\triangleright A_a^i = e^{-\l}A_a^i, 
\qquad \exp(\l W)\triangleright E^i_a = e^{\l}E^i_a,
\ee
where we used the canonical algebra satisfied by the connection and the triad.\footnote{Next to this kinematical dilatation, one could also consider a more `dynamical' dilatation, where the splitting of the Ashtekar-Barbero connection into a Levi-Civita part plus extrinsic curvature is taken into account. This would lead then to a nonlinear transformation, 
\be\notag
A'{}^i_a = \Omega^{-2} A^i_a + (1-\Omega^{-2})\G^i_a(E)+ 2 \eps^{ijk} E^i_a E^b_k \p_b \ln\Omega
\ee
which was considered for instance in \cite{Koslowski:2013ckb}.} In this section we only deal with a single copy of the cotangent bundle, so we do not need the label $i$ for the links, and we can safely use it for the $\su(2)$  indices.

Since the phase space $T^*\SU(2)$ has compact directions, there is no dilatation generator in the sense 
of \Ref{Wcont}. However, being ultimately interested in the continuum limit of the theory, it makes 
sense to consider a discretisation of $W$ acting on $T^*\SU(2)$. A simple choice is given by the Lie 
derivative of the group character, 
\be\label{defW}
W = 2 {\cal L}_X \chi^{(\f12)}(g) = -2\tr(g X),
\ee
which reduces to \Ref{Wcont} in the continuum limit, if we denote $g \simeq \Id + A$, $A := A^i_a \t_i \ell^a$, $X^i \simeq E_i$, $E_i:=E_i^a s_a$, with $\ell^a$ the coordinate tangent to the link at its source, and $s_a$ the coordinate area 2-form of its dual surface.
Let us study the transformations induced by \Ref{defW} on the holonomy-flux variables. At the infinitesimal level, we have
\be
\{W, g \} = \Id -\f{\tr g}2 g, \qquad \{W, X^i \} = \f12\tr g \, X^i - \eps^{ijk} X^j \tr(\t^k g).
\ee
To compute the finite action, it is easier to use to use the spinorial parametrization \Ref{T*SU2}. 
Interestingly, $W$ acts as a boost in mixing the spinors,
\be\label{Wspinors}
e^{\l W} \triangleright \vet{z^A}{\tl z^A} = \mat{\cosh\l/2}{\sinh\l/2}{\sinh\l/2}{\cosh\l/2}\vet{z^A}{\tl z^A}.
\ee
Using \Ref{Wspinors}, it is not hard to show that 
\begin{subequations}\label{Waction}
\begin{align}\label{gW}
& g_W := e^{\l W} \triangleright g = \f1{\cosh\l+\f{\tr g}2 \sinh\l}\left[ \cosh^2\f\l2 g + \sinh^2\f\l2 g^{-1} + \sinh\l \Id \right], 
\\
& X_W^i := e^{\l W} \triangleright X^i = X^i + \left( \sinh^2\f\l2+\f{X^2 \tr g}{2(X^2-X\cdot\tl X)} \sinh\l\right)(X^i-\tl X^i)
-\f{\tr (gX)}{X^2-X\cdot\tl X}  \sinh\l \eps^{ijk} X^j \tl X^k.
\end{align}\end{subequations}

The transformations are nonlinear,\footnote{The nonlinearity of the transformations \Ref{Waction} can be organised in powers of $\tr g$ and $\tr (gX)$, since 
$2(X^2 - X\cdot \tl X) = j^2\tr^2 g + 4 \tr^2(gX)$. It is intriguing that similar nonlinear structures with class-function factors $\tr g$ and $\tr (gX)$ appear in the construction of a phase space for curved simplices \cite{Bonzom:2014wva,Haggard:2015ima}, related in turn to Poisson-Lie groups and Hopf algebras. However, we do not know at this stage if $W$ has any specific role to play there.} as expected,
and can not be visualised as symplectic dilatations in the continuum sense \Ref{Wcont}.
Nonetheless, they have some interesting properties. 
First of all, they preserve the Poisson structure. To see that, let us rewrite them in a compact way as follows,
\be
g_W = S^\dagger g S^\dagger
, \qquad X_W := X_W^i\t_i =  {\cal N} (S X S^\dagger), 
\ee
where
\be
{\cal N} = \cosh \l +\f{\tr g}2 \sinh\l,\qquad
S = \f1{\sqrt{\cal N}}(\cosh \f\l2 \Id + \sinh\f\l2 g).
\ee
It is easy to show that $S\in \SU(2)$, and thus also $g_W$. Hence,
the transformation preserves the polarisation.
To verify preservation of the Poisson brackets, again it is easiest to use spinors: \Ref{Wspinors} is easily seen to preserve the spinorial brackets, and the area matching constraint is changed but only by a non-zero multiplicative factor, 
$e^{\l W} \triangleright C = {\cal N} C$. Hence, the symplectic reduction is still $T^*\SU(2)$.

Furthermore, it is easy to check that the transformations \Ref{Waction} have the right continuum limit: for $g \approx \Id + A$, 
we get
\begin{align}\label{Wcontlim}
& g_W \simeq \Id + e^{-\l} A, \qquad X^i_W \simeq e^\l X_i.
\end{align}
This means that the holonomy-flux `dilatation' operator captures in the continuum limit the properties of a connection-triad dilatation: rescaling of areas and volumes with preservation of angles, and rescaling of the connection. 
Furthermore, we will see below that in the same continuum limit, the covariant version of $W$ reproduces the linear shifts of $\Xi$ generated by $D$.

On the other hand, let us stress that unlike $D$, this action is \emph{not} generally compatible with the discrete geometric interpretation, since the closure condition is disrupted:
\be
e^{\l W} \triangleright \sum_{l\in n} \vec X_l = \sum_{l\in n} \vec X_{Wl} \neq 0,
\ee
a breaking that occurs because of the nonlinearity of the transformation.
At leading order in the continuum limit the closure defect can be approximated with $\sum_{l\in n} e^{\l_l}\vec X_{l} $, thus in this limit the symmetry is recovered at least for global transformations.\footnote{For graphs with $L\geq 3N$, it may also be possible to find non-trivial sets of $\l_l$ for which the closure defects vanish everywhere.}

The situation can be compared with Regge calculus. Since the fundamental variables are now lengths instead of areas and angles as in twisted geometries, local conformal transformations are defined rescaling the edge lengths at both ends  \cite{Rocek:1982tj},
$\d \ell_{xy} = (\l_x+\l_y)\ell_{xy}$. This ensures that the triangulation is not disrupted, however the volumes are changed by a scaling, and the angles are not invariant. Furthermore, it is not possible to extend this transformation to a a finite group action, because composition of two finite transformations generically breaks the triangle inequalities.
Nonetheless, this definition has useful applications to discrete Ricci flow \cite{Luo:fj,Glickenstein:2009kq} and quantum Regge calculus \cite{Marzuoli:2016qv}.

We conclude that while $W$ provides an interesting nonlinear definition of dilations in the compact phase space $T^*\SU(2)$, with a well-behaved continuum limit, its generic incompatibility with the closure constraint hinders direct applications to classical dynamical models on a fixed graph. On the other hand, it could intervene interestingly in situations where the closure constraint is relaxed, such as in coarse graining \cite{Livine:2013gna,RovelliHowManyQuanta} or in some spin foam models \cite{Bonzom:2009hw, AlexandrovSimplClosure}. 

\subsection{Transformations on twisted geometries}

For completeness, we report the action of $W$ on the various twisted geometry variables,
\begin{subequations}\label{Wtg}\begin{align}
& e^{\l W} \triangleright j = j \left( \cosh \l +\f{\tr g}2 \sinh\l \right), 
\\
& e^{\l W} \triangleright \z = \f{\cosh\f\l2 \sqrt{1+|\tl\z|^2}\, \z + \sinh\f\l2 \sqrt{1+|\z|^2}e^{\f i2\xi}\tl\z}{\cosh\f\l2 \sqrt{1+|\tl\z|^2} + \sinh\f\l2 \sqrt{1+|\z|^2}e^{\f i2\xi}}, 
\\
& e^{\l W} \triangleright \xi = 2\arctan \left[ \f{\sqrt{1+|\z|^2}\sqrt{1+|\tl\z|^2}\, \sin\xi/2}{\sqrt{1+|\z|^2}\sqrt{1+|\tl\z|^2}\, \cosh\l \cos\xi/2 + \Big(1+(|\z|^2+|\tl\z|^2)/2\Big) \sinh\l } \right]. 
\end{align}\end{subequations}
In particular the last one shows the strong departure from the simple action of $D$.
\footnote{Notice that for $\sinh^2\f\l2 = |\z|^{-2}$ the transformation on $\z$ reduces to a conformal 
transformation of $\C{\mathbb P}^1$ parametrized by $\tl \z$ and $\xi$. However, it is not possible to 
fix $\l$ so as to have a conformal transformation for both $\z$ and $\tl \z$.}  
In comparing the two operators, it is somewhat curious to point out that this complicated action stems 
from what looks like a rather minimal nonlinear operator: In the twisted geometries parametrization we can write
\be
W = -2j\p_\xi\tr g,
\ee
whereas as we have already seen on the reduced space that $D=\g j$.

On the gauge-invariant variables, we have
\be\label{Wbk}
e^{\l W_i} \triangleright \bk{i}{j} = \cosh\f\l2 \bk{i}{j} + \sinh\f\l2 \bk{\tl \imath}{j}, 
\ee
from which we deduce that 
\be
e^{\l_j W_j} \triangleright \xi^i_{jk} \neq e^{\l_k W_k} \triangleright \xi^i_{jk},
\ee
so a general transformation generated by $W$ will also disrupt the matching of shapes. 

We can also reproduce the continuum limit results. In the parametrization of twisted geometries, the continuum limit used in \Ref{Wcontlim} reads
$\tl\z\simeq\z+\d\tl\z$ and $ \xi\simeq\d\xi$, and $A=\d \tl n n^{-1} + \d\xi n\tau_3n^{-1}$.
In this limit, \Ref{Wtg} and \Ref{Wbk} reduce to 
\be
e^{\l W} \triangleright j \simeq e^{\l}j, \qquad 
e^{\l W} \triangleright \z \simeq \z + \f12 (1-e^{-\l}) \d\tl\z, \qquad
e^{\l W} \triangleright \xi \simeq e^{-\l}\d\xi, \qquad
e^{\l W_i} \triangleright \bk{i}{j} \simeq e^{\l}  \bk{i}{j}.
\ee
We recover in this way the preservation of angles and shape-matching conditions, as well as the rescaling of the part of the connection carried by $\xi$.

\subsection{Covariant action}
Finally, let us consider a covariant version of $W$, and study its action on the extrinsic geometry $\Xi$. At first sight, we could simply take 
\be
W' := -4\tr(h\Pi) = \pi\tl\om +\om\tl\pi. 
\ee
However, on shell of the simplicity constraints this mixes $g$ and $\hat g$:
\be
W' \approx -2(1-i\g)\left[\cosh\left((1+i\g)\f\Xi2 \right) \tr(gX) + \sinh\left((1+i\g)\f\Xi2 \right) \tr(\hat gX) \right].
\ee
The mixing can be easily avoided considering also 
\be
\hat W' := -4\tr(\hat h\Pi) \equiv \po\tr h = \pi\tl\om - \om\tl\pi,
\ee
and taking the following linear combination, 
\be
{\cal W} := \f{\cosh\left((1+i\g)\f\Xi2 \right) W' - \sinh\left((1+i\g)\f\Xi2 \right) \hat W'}{(1-i\g) \cosh\big[(1+i\g)\Xi \big]}
\equiv \f{e^{-(1+i\g)\f\Xi2} \pi\tl\om + e^{(1+i\g)\f\Xi2} \om\tl\pi }{(1-i\g) \cosh\big[(1+i\g)\Xi \big]}
\approx  -2\tr(gX) \equiv W.
\ee
Equipped with this covariant version of the generator of connection-triad symplectic dilatations, we can evaluate its action on $\Xi$, for which we obtain 
\begin{align}\notag
\{{\cal W}, \Xi \} &= \f{1}{(1-i\g) \cosh\big[(1+i\g)\Xi \big]} \left(e^{-(1+i\g)\f\Xi2} \f{\bk{\om}{\tl\om}}{||\om||^2} + e^{(1+i\g)\f\Xi2} \f{\bk{\tl\om}{\om}}{||\tl\om||^2}\right) \\
& \approx \f2{1-i\g}\sqrt{\f{1-N\cdot \tl N}2 } \left[  
\cos\left(\f\xi2+\arg(1+\tl \z\bar\z)\right) - i \sin\left(\f\xi2+\arg(1+\tl \z\bar\z)\right) \tanh\left((1+i\g)\Xi \right) \right].
\end{align}
Expanding at first order in the continuum limit $h=\Id+o(A)$, we get
\be
\{{\cal W}, \Xi\} \simeq \f2{1-i\g}.
\ee
Up to a numerical factor, which is due to the fact that we started with $\Pi$ instead of the rotation generator $L=\Pi^i+\bar\Pi^i$, we have reproduced the translation transformation generated by $D$.

To summarise, comparing the action of $\cal W$ with $D$, we see that at leading order in the continuum limit \Ref{Wcontlim}, the action of $\cal W$ changes areas and volumes, preserves the 2d and 3d angles, \emph{and} reproduces the shift in extrinsic curvature caused by $D$. This is an interesting property, however as already stated, care is needed in the use of $\cal W$ as the shifts in the areas are in general incompatible with the closure constraint.

\section{Conclusions}

The main goal of the paper was to study the action of the group of
conformal isometries of Minkowski space on the phase space of loop
quantum gravity on a fixed graph, in its description in terms of
twisted geometries. To do so, we used the twistorial parametrization,
and the Hamiltonian action of $\SU(2,2)$ on twistor space. We
showed that translations and special conformal transformations are not
compatible with the various constraints reducing twistor space to
twisted geometries, but that the dilatation generator is. The origin
of this compatibility is to be found in a previously unstudied
symmetry of embeddings of $\SL(2,\C)$ in twistor space, or
equivalently of $T^*\SU(2)$ in $\C^4$. The associated orbits have an
interesting geometric meaning. They change the
extrinsic geometry by a linear shift, described by boosts among adjacent polyhedra, as
well as the embedding of the SU(2) holonomy in the covariant phase
space, an embedding handled at the continuum level by the
Barbero-Immirzi canonical transformation and imposition of secondary
constraints.

In doing so, we highlighted the role of the diagonal simplicity and
area matching in breaking full conformal invariance, and discussed the
geometric meaning of the various constraints from the viewpoint of
twistor theory, showing an isomorphism with null twistors, and how the
simplicity constraints identify a space-like plane that is then used
for the geometric reconstruction. Our analysis can be extended in many
directions. First, investigating the dynamical applications of $D$,
either in a conventional setting with flat 4-simplices, see also
\cite{FreidelConfLambda} on this, or in an alternative setting with
curved 4-simplices, as advocated for instance in \cite{Bahr:2011uj}.
In the latter case, while the twistorial description of the conformal
group immediately applies, the usual phase space description in terms
of twisted geometries has to be changed, presumably along the lines
investigated in \cite{Bonzom:2014wva,Haggard:2015ima}.
Similarly, the geometric meaning and possible application of $W$
deserves further study. Since symplectic dilatations generate squeezed
coherent states for a particle on a line, it would be interesting to
see whether $W$ can be used to construct interesting squeezed spin
networks, a topic of  recent interest in the community
\cite{Bianchi:2015fra, Hamma:2015xla,Feller:2015yta}.

Finally, we hope to come back in future research to the more formal
aspect of our motivations, and use twistor methods to study conformal
spin networks and their applications to loop quantum
gravity.\footnote{Here we have in mind $\SU(2,2)$ spin networks, as
mentioned in the introduction. A relation between spin networks and
the infinite dimensional group of conformal transformations in 2
dimensions has been hinted at recently in
\cite{Ghosh:2014rra,Freidel:2015gpa,Bonzom:2015ans}.}

\subsection*{Acknowledgments}
We thank Bianca Dittrich, Tim Koslowski, Laurent Freidel, Etera Livine and Wolfgang Wieland for discussions, and Laurent Freidel for sharing a draft of \cite{FreidelConfLambda}. ML acknowledges grant no. 266101 by 
the Academy of Finland.

\appendix

\subsection*{A. Algebra Conventions}
We take the generators of $\su(2,2)$ to satisfy
\be\label{MM}
\{M^{ab}, M^{cd}\} = \eta^{ac} M^{bd} -\eta^{ad} M^{bc} + \eta^{bd} M^{ac} - \eta^{bc} M^{ad},
\ee
with $a=0$ to 5, and $\eta^{ab}={\rm diag}(-++++-)$. We further fix conventions  $\eps_{012345} = \eps^{012345}=\eps^{0123} =1$, $\eps^{IJKL}=\eps^{IJKL45}$.
To highlight the various subalgebras of $\su(2,2)$, we introduce the following notation, 
\begin{align}
J^{IJ}=M^{IJ}, \qquad P^I=\f1{\sqrt{2}}(M^{I5}+M^{I4}), \qquad C^I=\f1{\sqrt{2}}(M^{I5}-M^{I4}), \qquad D=-M^{45},
\end{align}
where $I=0$ to 3. The $J$ generate the Lorentz subalgebra, and can be further decomposed as 
\be
K^i = L^{0i}, \quad L^i = -\f12\eps^i{}_{jk}L^{jk}, \quad \Pi^i = i J^{0i}_- = \f12(L^i+i K^i).
\ee
In terms of this decomposition, the Poisson brackets \Ref{MM} read
\begin{align}
& \{L^i,L^j\} = -\eps^{ijk} L^k, && \{L^i,K^j\} = -\eps^{ijk} K^k, && \{K^i,K^j\} = \eps^{ijk} L^k, \notag \\
& \{P^I,J^{JK}\} = -\eta^{IJ} P^K + \eta^{IK} P^J, && \{C^I,J^{JK}\} = - \eta^{IJ} C^K + \eta^{IK} C^J,  \label{confalg} \\
& \{C^I, P^J\} = \eta^{IJ} D -  J^{IJ}, && \{D, P^I \} = - P^I, && \{D, C^I\} = C^I. \notag
\end{align}

In the main text we refer to the three Casimir invariants of $\su(2,2)$. These are given by 
\begin{align}
\mathcal{C}^{(2)} &= M_{ab} M^{ab} = 2C_1 - 2D^2 - 4 P\cdot C, \\
\mathcal{C}^{(3)} &= \f18 \eps_{abcdef} M^{ab} M^{cd} M^{ef}  \label{Cas3}
=3DC_2 - 6 W\cdot C, \\ \notag
\mathcal{C}^{(4)} &= M_{ab} M^{bc} M_{cd} M^{da}  \\
& = 2C_1{}^2+C_2{}^2 + 2D^4 +8 J_{IJ} J^{JK} P^I C_K + 8 D J_{IJ} P^I C^J + 8D^2 P\cdot C + 4 P^2 C^2+4 (P\cdot C)^2,  
\end{align}
where 
\begin{align}
C_1 = \f12J_{IJ}J^{IJ} = L^2-K^2 = 2(\Pi^2+\bar \Pi^2), \qquad C_2 = \f12(\star J)_{IJ}J^{IJ} = 2 K\cdot L = 2i(\Pi^2-\bar\Pi^2)
\end{align}
are the Lorentz Casimirs, and 
\be
W_I = \f12 \eps_{IJKL} M^{JK} P^L 
\ee
is the Pauli-Lubanski vector, which satisfies $W_I = -s P_I$ for a massless particle.
In deriving ${C}^{(4)}$, 
we used the identities
\begin{align}
& J_{IJ} J^{JK} P^I C_K = -\f12\eps_{IJKL} J^{IJ} W^K C^L - \f12 J_{IJ} J^{IJ} P\cdot C, \qquad
J_{IJ}J^{JK}J_{KL}J^{LI} =  2C_1{}^2+C_2{}^2. 
\end{align}

\subsection*{B. Spinorial notation}
\noindent

Throughout the paper, we use regularly Penrose's abstract index convention and spinor formalism. 
The key difference between our conventions and those more commonly used in the twistor literature (e.g. \cite{PenroseRindler2}) is the metric signature, which we take to be mostly plus. Then, the abstract index map $I=A\dot A$ can be realized in terms of \emph{anti}-hermitian matrices. That is,
\be
X^{A\dot A} = \f{i}{\sqrt{2}} \s_I^{A\dot A} X^{I} =  \f{i}{\sqrt{2}} \mat{X^0+X^3}{X^1-iX^2}{X^1+iX^2}{X^0-X^3}.
\ee
Spinorial indices are raised and lowered with the antisymmetric $\eps_{AB}$ symbol, 
$\om_A=\om^B\eps_{BA}$, with $\eps_{01}=1$. We further use the following ket notation to eliminate explicit indices in most formulas:
\begin{align}
& \ket{\om}=\om^A, \qquad \bra{\om} = \ket{\om}^\dagger = \d_{A\dot A}\bar\om^{\dot A}, \qquad \|\om\|^2=\bra{\om}\om\ra,
\\
& |\om] := -\eps \ket{\bar\om} = -\d^{A\dot B} {\epsilon}_{\dot B\dot A} \bar\om^{\dot A}, \qquad [\om| = \om_A = \om^B \eps_{BA}, 
\qquad [\pi\ket{\omega} = \eps_{AB} \pi^A \om^B.\label{dctnry}
\end{align}
Finally, we take self-dual projectors
$P_\pm{}^{IJ}_{KL} = \f12 (\Id\mp i \star),$ satisfying
$\star P_\pm = \pm i P_\pm.$ Following Penrose, we call self-dual (or right-handed) the positive eigenspace, and map it to dotted indices, and antiself-dual (or left-handed) the negative one, and map it to undotted indices. Accordingly, for a bivector we have
\be
F^{IJ} = P_-{}^{IJ}_{KL} F^{KL}+P_+{}^{IJ}_{KL} F^{KL}
= F^{AB\dot A\dot B} = \phi^{AB}\eps^{\dot A\dot B} + \bar\phi^{\dot A\dot B}\eps^{AB} \in (1,0)\oplus(0,1),
\ee
with left-handed part $\phi^{AB}=-F^{0i}_{-}\s^{AB}_i$, and right-handed part its complex conjugate.

\providecommand{\href}[2]{#2}\begingroup\raggedright\endgroup

\end{document}